\documentclass[twocolumn,twocolappendix]{aastex631}

\newcommand{\msun}{M_\odot}

\newcommand{\kms}{km~s$^{-1}$}

\usepackage{CJK}
\usepackage{acronym}
\usepackage{xcolor}
\usepackage{ulem}
\usepackage{amsmath}
\usepackage{bm}

\acrodef{ns}[NS]{neutron star}
\acrodef{sn}[SN]{supernova}
\acrodefplural{sn}[SNe]{supernovae}
\acrodef{tzo}[T\.ZO]{Thorne-\.Zytkow object}
\acrodef{grb}[GRB]{gamma-ray burst}
\usepackage{gensymb}

\begin{document}
\begin{CJK*}{UTF8}{ipxm}

\title{Neutron star kicks plus rockets as a mechanism for forming wide low-eccentricity neutron star binaries}

\author[0000-0002-8032-8174]{Ryosuke Hirai (平井遼介)}
\affiliation{Astrophysical Big Bang Laboratory (ABBL), Cluster for Pioneering Research, RIKEN, Wako, Saitama 351-0198, Japan}
\affiliation{School of Physics and Astronomy, Monash University, Clayton, VIC3800, Australia}
\affiliation{OzGrav: The ARC Centre of Excellence for Gravitational Wave Discovery, Clayton, VIC3800, Australia}

\author[0000-0002-8338-9677]{Philipp Podsiadlowski}
\affiliation{University of Oxford, St Edmund Hall, Oxford OX1 4AR, UK}

\author[0000-0002-3684-1325]{Alexander Heger}
\affiliation{School of Physics and Astronomy, Monash University, Clayton, VIC3800, Australia}
\affiliation{Argelander-Institut für Astronomie, University of Bonn, Auf dem H{\"u}gel 71, 53121 Bonn,	Germany}

\author[0000-0002-7205-6367]{Hiroki Nagakura (長倉洋樹)}
\affiliation{Division of Science, National Astronomical Observatory of Japan, 2-21-1 Osawa, Mitaka, Tokyo 181-8588, Japan}

\begin{abstract}
Recent neutron star surface observations corroborate a long-standing theory that neutron stars may be accelerated over extended periods after their birth. We analyze how these prolonged rocket-like accelerations, combined with rapid birth kicks, impact binary orbits. We find that even a small contribution of rocket kicks combined with instantaneous natal kicks can allow binaries to reach period--eccentricity combinations unattainable in standard binary evolution models. We propose these kick + rocket combinations as a new channel to form wide low-eccentricity neutron star binaries such as Gaia NS1, as well as inducing stellar mergers months to years after a supernova to cause peculiar high-energy transients. 
\end{abstract}

\section{Introduction}

Neutron stars (NSs)\acused{ns}, the compact remnants of massive stars, are known to be born with high spatial velocities of a few$\times100$~km~s$^{-1}$ \citep[]{lyn94}. In the current paradigm, this velocity is given at birth due to asymmetries in the \ac{sn} explosion, driven by hydrodynamical instabilities \citep[]{bur95,jan17} and asymmetric neutrino emission \citep[]{arr99,tam14,nag19,col22,bur24}. Since this velocity is related to the core-collapse \ac{sn} explosion mechanism itself, the acceleration occurs on dynamical timescales of $\tau_\mathrm{kick}\lesssim10$~s and the resulting velocity is often called a ``natal kick''.

On the other hand, some kick mechanisms have been proposed in the past that involve \textit{post-natal} processes. A famous example is the ``electromagnetic rocket'' model, where the new-born \ac{ns} is propelled along the spin axis by asymmetric spin-down radiation from off-centred or higher-order magnetic fields \citep[]{har75,lai01,koj11,pet19,igo23}. 
Recently, the NICER X-ray telescope revealed the non-antipodal orientation of hot spots on a neutron star surface, strongly supporting the existence of such higher-order magnetic fields \citep[]{mil19,ril19,kal21} and therefore it is very likely that the electromagnetic rocket mechanism contributes to the \ac{ns} kick to some degree. The total magnitude of the velocity achieved through this mechanism scales as
\begin{equation}
 v_\mathrm{roc}\approx260~\mathrm{km~s}^{-1}\frac{\bar{\epsilon}}{0.1}\left(\frac{R}{10~\mathrm{km}}\right)^2\left(\frac{P_\mathrm{spin,ini}}{1~\mathrm{ms}}\right)^{-2},
\end{equation}
where $\bar{\epsilon}$ expresses the degree of asymmetry of the outgoing Poynting flux, $R$ and $P_\mathrm{spin,ini}$ are the \ac{ns} radius and birth spin period respectively \citep[]{lai01}. In the most optimal magnetic field configuration, this can reach up to $v_\mathrm{roc}\sim1\mathord,400$~km~s$^{-1}$\footnote{It should be noted that the strength of the magnetic field only influences the acceleration timescale in this model. The final velocity is determined by the reservoir of rotational energy and asymmetry of the radiation, hence the field strength is irrelevant.}. Another post-natal model is the so-called ``neutrino rocket'' mechanism, where neutrinos are asymmetrically emitted via neutrino cyclotron radiation from neutrons circulating superfluid vortices in the \ac{ns} interiors \citep[]{pen82,pen04,li22}. Under favourable assumptions, this mechanism can accelerate the \ac{ns} to velocities reaching $v_\mathrm{roc}\lesssim1,000$~\kms. Although some difficulties have been pointed out for these ``rocket'' mechanisms, such as the requirement of a high birth spin ($P_\mathrm{spin}\sim1$~ms) that is in contradiction with observationally inferred values \citep[]{lai01,jan22}, these models have not been firmly ruled out \citep[]{aga23}. In particular, there are no arguments against the rocket mechanisms making minor contributions to the total kick.

A distinct property of the rocket mechanisms is its long acceleration timescale. Since it is driven by spin-down, the acceleration timescale can be $\tau_\mathrm{roc}\sim0.1$--$100$~yr ($\gg\tau_\mathrm{kick}$). Whereas this is still negligible compared to the lifetime of pulsars ($\sim10^{7-8}$~yr), meaning it will be difficult to detect ongoing acceleration, it is longer than typical orbital periods of binary systems containing \acp{ns}. Traditionally, \ac{ns} kicks in the context of binary evolution have been treated as an impulsive process, in which the post-\ac{sn} orbit can be analytically derived. The resulting orbital properties depend on the magnitude and direction of the kick, but in most cases it causes the orbit to become wider and eccentric or disrupt it completely. However, when the kick cannot be treated impulsively but instead the acceleration timescale is comparable to, or longer than, the orbital period, these impulsive solutions no longer hold. Instead, we need to consider the opposite extreme where the \ac{ns} acceleration is a small perturbation compared to the orbital acceleration.

The two-body problem with a constant acceleration on one or both of the bodies is sometimes called the ``accelerated Kepler problem'' \citep[]{nam07,lan11}, which is analogous to the Stark effect in quantum mechanics. Due to the secular nature, the orbit can be altered in ways that are qualitatively different from instantaneous kicks. It has been studied broadly in the context of planetary sciences, artificial satellites and white dwarf binaries \citep{nam05,nam07,hey07,lan11}. The impact on \ac{ns} binaries, however, has not been well quantified and the existing studies only deal with special cases where the rocket mechanisms act on initially circular orbits \citep[]{har75,pet19}. 

In this paper, we explore the implications of rocket-like acceleration in tandem with classical impulsive kicks, in particular, how it influences our understanding of the formation channels for various \ac{ns} binaries. 
One of the main consequences is that it provides a channel to form the wide low-eccentricity \ac{ns} binary Gaia NS1 \citep{elb24a} and other similar systems \citep[]{hin06,hin24,elb24b,nag24} that cannot be understood in standard binary evolution models. This may provide direct testable evidence for the rocket mechanisms in \ac{ns} binaries. 

This paper is structured as follows. In Section~\ref{sec:AKP}, we introduce the secular accelerated Kepler problem and derive the analytic solution. The combined effect of rapid \ac{ns} kicks and the rocket effect are demonstrated in Section~\ref{sec:combination}, and its astrophysical implications are discussed in Section~\ref{sec:implications}. We summarize our conclusions in Section~\ref{sec:conclusion}.

\section{The accelerated Kepler problem}\label{sec:AKP}

In this section we introduce the accelerated Kepler problem (or classical Stark problem), which is a two-body problem with a fixed acceleration on one of the bodies. The equation of motion of our two-body problem is
\begin{align}
    m_1\ddot{\bm{r}}_1&=\bm{F}_{12}+m_1\bm{a}_\mathrm{roc},\label{eq:eom1}\\
    m_2\ddot{\bm{r}}_2&=\bm{F}_{21},\label{eq:eom2}
\end{align}
where $m_i$ are the masses, $\bm{r}_i$ are the position vectors where subscripts $i=1,2$ denote the primary and secondary, and $\bm{a}_\mathrm{roc}$ is the rocket acceleration acting on the new-born \ac{ns}. $\bm{F}_{12}=-\bm{F}_{21}$ is the gravitational force between the two stars. Dots denote time derivatives. To solve for the orbit, we separate the equations into the centre of mass and relative motion by defining
\begin{align}
 &\bm{r}_\mathrm{com}\equiv\frac{m_1\bm{r}_1+m_2\bm{r}_2}{m_1+m_2},\\
 &\bm{r}\equiv\bm{r}_1-\bm{r}_2.
\end{align}
We rewrite the gravitational force as
\begin{equation}
 \bm{F}_g\equiv\bm{F}_{12}=-\bm{F}_{21}=-\frac{Gm_1m_2}{r^2}\hat{\bm{r}}=-\frac{k}{r^2}\hat{\bm{r}},
\end{equation}
where we have defined $k\equiv Gm_1m_2$ and the unit vector $\hat{\bm{r}}=\bm{r}/|\bm{r}|$. The equation of motion for the centre of mass and relative position can hence be expressed as
\begin{align}
 &\ddot{\bm{r}}_\mathrm{com}=\frac{m_1\bm{a}_\mathrm{roc}}{m_1+m_2},\label{eq:eom_com}\\
 &\ddot{\bm{r}}=\frac{\bm{F}_g}{\mu}+\bm{a}_\mathrm{roc},\label{eq:relative_eom2}
\end{align}
where $\mu\equiv m_1m_2/(m_1+m_2)$ is the reduced mass. The same expression for Eq.~(\ref{eq:relative_eom2}) will hold even for cases where both stars have separate accelerations, by replacing $\bm{a}_\mathrm{roc}$ with the difference of the two accelerations.

\subsection{Deriving the analytical solution}\label{sec:analytics}
The problem has been known to be integrable for more than two centuries, but it was only in the work by \citet{lan11} that the general closed-form solution to this problem was shown using elliptical coordinates. Under the secular approximation ($|\bm{a}_\mathrm{roc}|\ll|\bm{F}_\mathrm{g}/\mu|$), the semi-major axis (or orbital period) is preserved whilst the orbit is torqued by the rocket acceleration \citep{hey07}. The general behaviour is that the orbital plane librates around the direction of the rocket and the eccentricity oscillates up and down, while the binary as a whole is also accelerated. In this section, we derive an alternate analytical solution to the secular accelerated Kepler problem in vector form. Our much simpler form allows for an intuitive understanding and makes it suitable for statistical investigation. 

Our solution can be summarized as follows. The energy and angular momentum of the orbit are defined as
\begin{align}
  E&\equiv\frac12\mu\dot{r}^2-\frac{k}{r}=-\frac{k}{2a},\label{eq:orbital_energy}\\
  \bm{L}&\equiv\mu \bm{r}\times\dot{\bm{r}},\label{eq:orbital_AM}
\end{align}
where $a$ is the semi-major axis.
The energy gives an average orbital velocity
\begin{equation}
 \bar{v}_\mathrm{orb}=\sqrt{-\frac{2E}{\mu}}=\sqrt{\frac{k}{\mu a}}.
\end{equation}
Let us define a dimensionless angular momentum vector $\bm{l}\equiv \bm{L}/L_0$ where $L_0\equiv k/\bar{v}_\mathrm{orb}$ is the angular momentum of a circular orbit ($e=0$) with the same energy. The magnitude of this vector is $|\bm{l}|=\sqrt{1-e^2}$.
By taking the time derivative of the normalized angular momentum and using Eq.~(\ref{eq:relative_eom2}), the torque on the orbit by the rocket acceleration can be expressed as
\begin{equation}
 \dot{\bm{l}}=\frac{\mu}{L_0}\bm{r}\times\bm{a}_\mathrm{roc}.
\end{equation}
Notice that $\dot{\bm{l}}\cdot\bm{a}_\mathrm{roc}=0$, so the angular momentum in the direction of the rocket is strictly constant over time, whereas the transverse components are torqued by the rocket acceleration. Making the secular approximation, this becomes
\begin{align}
 \left<\dot{\bm{l}}\right>&=\frac{\mu}{L_0}\left<\bm{r}\right>\times\bm{a}_\mathrm{roc},\nonumber\\
 &=-\frac{3}{2\bar{v}_\mathrm{orb}}\bm{e}\times\bm{a}_\mathrm{roc},\label{eq:ldot}
\end{align}
where the angle brackets represent quantities averaged over an orbit 
\begin{equation}
    \left<\bm{x}\right>\equiv\frac{1}{P_\mathrm{orb}}\int_0^{P_\mathrm{orb}}\bm{x}(t)dt,
\end{equation}
and the eccentricity vector (or Laplace-Runge-Lenz vector) is defined as
\begin{equation}
 \bm{e}\equiv\frac{1}{k}\dot{\bm{r}}\times\bm{L}-\hat{\bm{r}}.
\end{equation}
We have also used that $\left<\bm{r}\right>=-3a\bm{e}/2$ and $k/(\mu a)=\bar{v}_\mathrm{orb}^2$.

Similarly, if we take the time derivative of the eccentricity vector, we obtain
\begin{equation}
 \dot{\bm{e}}=\frac{1}{k}\left(\bm{a}_\mathrm{roc}\times\bm{L}+\dot{\bm{r}}\times\dot{\bm{L}}\right),
\end{equation}
and the orbit-averaged value is
\begin{align}
 \left<\dot{\bm{e}}\right>&=\frac{1}{\bar{v}_\mathrm{orb}}\left(\bm{a}_\mathrm{roc}\times\bm{l}+\left<\dot{\bm{r}}\times\dot{\bm{l}}\right>\right),\nonumber\\
 &=-\frac{3}{2\bar{v}_\mathrm{orb}}\bm{l}\times\bm{a}_\mathrm{roc}.\label{eq:edot}
\end{align}
Note here that $\left<\dot{\bm{e}}\right>\cdot\bm{a}_\mathrm{roc}=0$, so the component of the eccentricity vector parallel to the rocket is constant under the secular approximation.

We now define the following two vectors to aid our understanding:
\begin{equation}
 \bm{h}_\pm\equiv\bm{l}\pm\bm{e},
\end{equation}
There are several properties to these vectors. First, $|\bm{h}_\pm|=1$, which is trivial from $|\bm{e}|=e$, $\bm{e}\cdot\bm{l}=0$ and $|\bm{l}|=\sqrt{1-e^2}$. Secondly, $\left<\dot{\bm{h}_\pm}\right>\cdot\bm{a}_\mathrm{roc}=0$, because $\dot{\bm{l}}\cdot\bm{a}_\mathrm{roc}=\left<\dot{\bm{e}}\right>\cdot\bm{a}_\mathrm{roc}=0$, meaning that the component of the $\bm{h}_+$ and $\bm{h}_-$ vectors parallel to the rocket are constant over time. Using Eqs.~(\ref{eq:ldot}) \& (\ref{eq:edot}), the secular evolution of these vectors can be expressed as
\begin{equation}
 \left<\dot{\bm{h}_\pm}\right>=\mp\frac{3}{2\bar{v}_\mathrm{orb}}\bm{h}_\pm\times\bm{a}_\mathrm{roc}.
\end{equation}
This implies that as the \ac{ns} is accelerated by the rocket mechanism, both $\bm{h}_+$ and $\bm{h}_-$ precess about the direction of the thrust $\bm{a}_\mathrm{roc}$ in opposite circular motions. Let us take the direction of the rocket $\bm{a}_\mathrm{roc}$ as the $z$ axis. If the cumulative rocket velocity up to some time $t$ is
\begin{equation}
 \Delta v_\mathrm{roc}(t)\equiv\int_0^t a_\mathrm{roc}dt, \label{eq:delta_vroc_def}
\end{equation}
 the post-rocket state can be described by 
\begin{align}
 \bm{h}_\pm'&=R_z\left(\pm\frac{3\Delta v_\mathrm{roc}}{2\bar{v}_\mathrm{orb}}\right)\bm{h}_\pm,\label{eq:hevo}
\end{align}
where $R_z(\theta)$ is a rotation matrix about the $z$-axis for an angle $\theta$. We can see that the two vectors will circulate around the $z$-axis in opposite directions. The post-rocket orbit can then be completely described by
\begin{align}
 \bm{l}'&=\frac12\left(\bm{h}_+'+\bm{h}_-'\right),\label{eq:postSN_lorb}\\
 \bm{e}'&=\frac12\left(\bm{h}_+'-\bm{h}_-'\right).\label{eq:postSN_ecc}
\end{align}
The post-rocket eccentricity is simply $e'=|\bm{e}'|$.

Although we have assumed that $\bm{a}_\mathrm{roc}$ is constant, our above solution still holds with variable acceleration magnitudes as long as it does not change significantly over an orbital period. In the rocket mechanism, the magnitude should slowly decline over time as the \ac{ns} spins down. Eventually, $\bm{a}_\mathrm{roc}$ should become negligible and the eccentricity oscillation would cease. The asymptotic value of the accumulated velocity $\Delta v_\mathrm{roc}$ determines the final eccentricity.

There are some complications if the rocket direction (or the spin axis) precesses. Our solution would still likely hold if the precession timescale is much shorter than the orbital period. On average, the rocket acceleration should be pointed along the axis of precession with a slightly lower magnitude, and the orbital evolution can be computed using this averaged acceleration. When the precession timescale is comparable to or longer than the orbital period, the above solution no longer holds. It has been shown in previous studies that in such cases the precessing rocket acceleration can lead to resonant eccentricity excitations \citep[]{nam05,nam07}. For even longer precession timescales, the rocket acceleration ceases before the spin direction changes appreciably, so again our above solution may be applicable.

\subsection{Numerical validation}

\begin{figure}
 \centering
 \includegraphics[width=\linewidth]{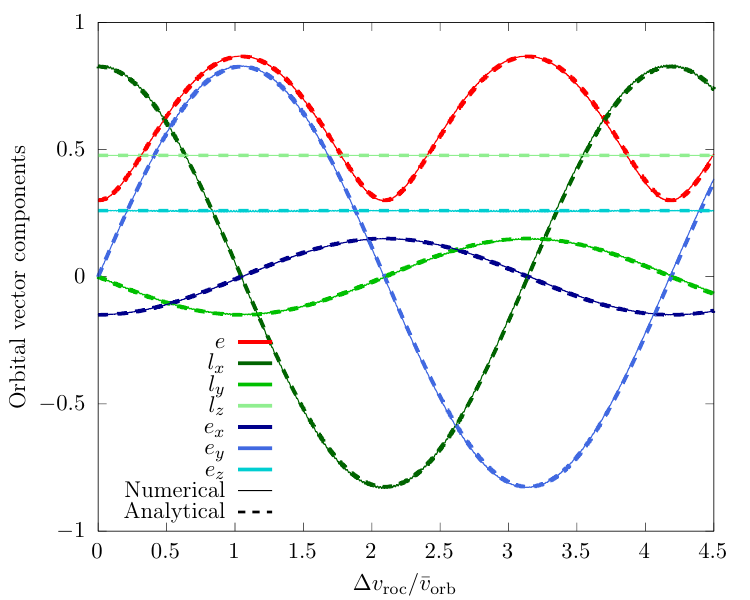}
 \caption{Example evolution of orbital vector components. The horizontal axis is the accumulated rocket velocity normalized by the average orbital velocity, which can be regarded as a measure of time (see Eq.~(\ref{eq:delta_vroc_def})). Results from direct numerical integration of Eqs.~(\ref{eq:eom1})--(\ref{eq:eom2}) are shown as thin solid curves and the analytical predictions are presented as thick dashed curves. The numerical model was evolved with an acceleration magnitude of $|\bm{a}_\mathrm{roc}|=\bar{v}_\mathrm{orb}/(50~P_\mathrm{orb})$. \label{fig:elements}}
\end{figure}

\begin{figure}
 \centering
 \includegraphics[width=\linewidth]{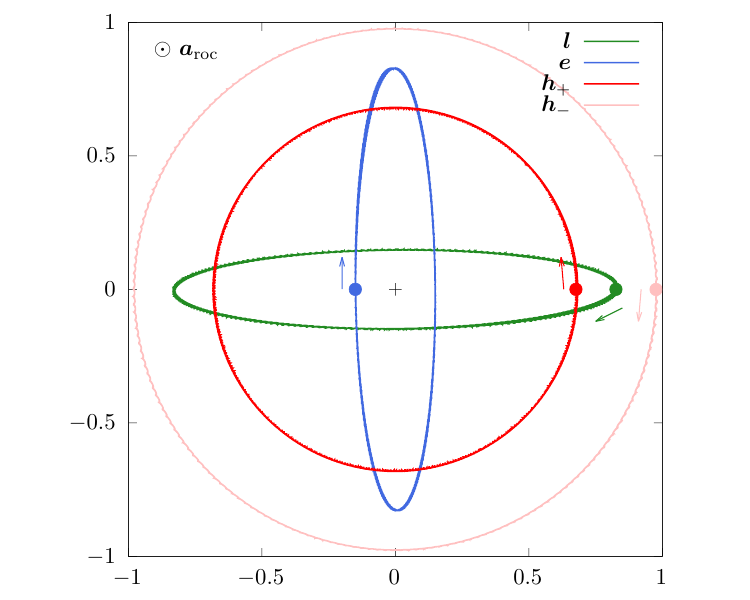}
 \caption{Trajectory of the orbital state vectors projected on the $xy$-plane. Circles indicate the initial values and the arrows point the direction of evolution. Parameters for the orbit are the same as in Figure~\ref{fig:elements}.\label{fig:xy_projections}}
\end{figure}

We verify that our analytic solution describes the true solution well by comparing against direct numerical integrations of the accelerated Kepler problem.
Figure~\ref{fig:elements} compares an example of the evolution of individual components of $\bm{l}$ and $\bm{e}$ based on Eqs.~(\ref{eq:postSN_lorb})--(\ref{eq:postSN_ecc}) (dashed curves) against direct numerical integration of Eqs.~(\ref{eq:eom1})--(\ref{eq:eom2}) (solid curves) with a low acceleration magnitude ($|\bm{a}_\mathrm{roc}|=\bar{v}_\mathrm{orb}/(50~P_\mathrm{orb})$). We start the integration from a binary with an eccentricity $e=0.3$ and the rocket direction is tilted from $\bm{l}$ by $60^\circ$ in the $x$ direction. We can see that the $xy$ components of the vectors $\bm{l}$ and $\bm{e}$ both oscillate sinusoidally with respect to the accumulated rocket velocity $\Delta v_\mathrm{roc}$. We show the trajectory of the vectors $\bm{l}$ and $\bm{e}$ projected on the $xy$-plane in Figure~\ref{fig:xy_projections}. Both vectors follow elliptic trajectories around the $z$-axis with the same semimajor and semiminor axes. When converted to the $\bm{h}_\pm$ vectors, we can see they follow circular trajectories moving in opposite directions. As a result of these circulations, the eccentricity oscillates, although not as a sine curve. The analytic solution is in perfect agreement with the numerical result.

\begin{figure}
 \includegraphics[width=\linewidth]{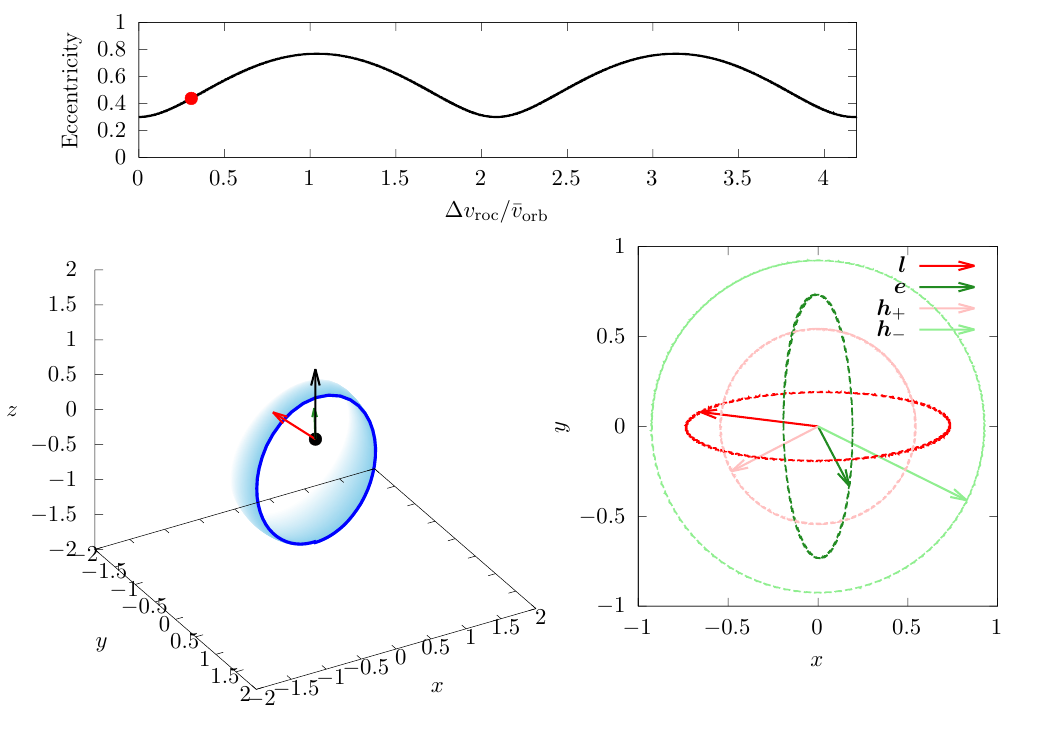}
 \caption{Animation of the time evolution of the orbit (bottom left panel), eccentricity (top panel) and projected orbital vectors (bottom right panel). In the bottom left panel, the coordinate origin is fixed on the \ac{ns} and the black arrow shows the direction of the rocket acceleration. The red and green arrows indicate the dimensionless angular momentum and eccentricity vectors. The blue ellipse shows the orbital trajectory of the companion. The animation shows the time evolution from $\Delta v_\mathrm{roc}/\bar{v}_\mathrm{orb}=0$ to $\Delta v_\mathrm{roc}/\bar{v}_\mathrm{orb}=4\pi/3$. (The animation of this figure is available online.)\label{fig:movie}}
\end{figure}

To aid the visual understanding of how the orbit evolves, we provide an animation of the orbital evolution along with how the eccentricity and projected orbital vectors evolve in Figure~\ref{fig:movie}. From the bottom left panel it is clear how the orbital plane librates while changing its eccentricity.

The amplitude of the eccentricity oscillation depends on the initial eccentricity and rocket angle. For example, when the initial orbit is circular and the rocket is aligned with the orbital angular momentum ($\bm{h}_+=\bm{h}_-=\hat{\bm{a}}_\mathrm{roc}$), there is no orbital evolution. On the other hand, if the initial orbit is circular and the rocket is perpendicular to the orbital angular momentum, the eccentricity can reach unity $e=1$.

\begin{figure}
 \centering
 \includegraphics[width=\linewidth]{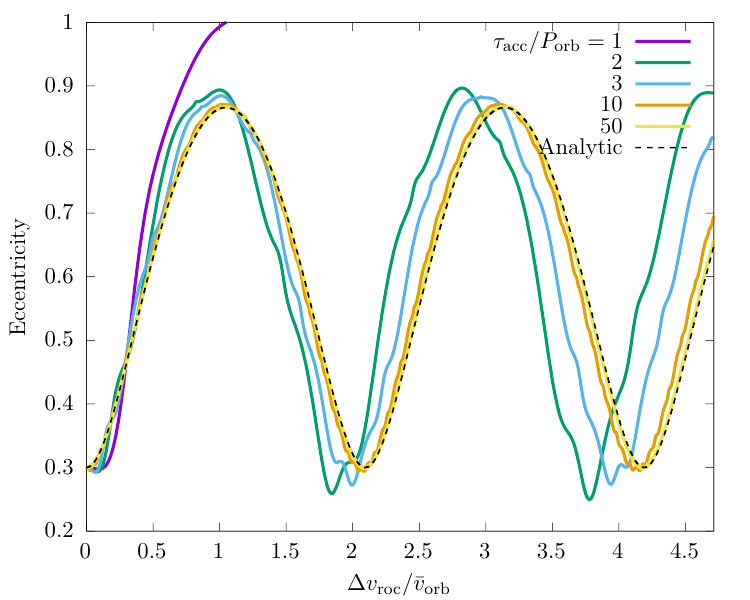}
 \caption{Evolution of the orbital eccentricity as a function of the accumulated rocket velocity. Solid curves show results from direct numerical integration with various acceleration timescales $\tau_\mathrm{acc}$. The black dashed curve is our analytical solution. Orbital parameters are the same as in Figure~\ref{fig:elements}.\label{fig:ecc_evo_numerical}}
\end{figure}

To assess when the secular approximation breaks down, we perform numerical integrations with various acceleration timescales $\tau_\mathrm{acc}\equiv\bar{v}_\mathrm{orb}/|\bm{a}_\mathrm{roc}|$. The results are shown in Figure~\ref{fig:ecc_evo_numerical}. As we demonstrated above, when the acceleration timescale is taken to be sufficiently longer than the orbital period $\tau_\mathrm{acc}/P_\mathrm{orb}\gtrsim10$, the evolution of the eccentricity closely follows our secular solution (dashed curve). At $\tau_\mathrm{acc}/P_\mathrm{orb}\lesssim1$, the secular approximation completely breaks down and hence the binary is dissociated within a single orbit. For the intermediate acceleration timescales where it is larger than but comparable to the orbital period ($2\lesssim\tau_\mathrm{acc}/P_\mathrm{orb}\lesssim10$), the numerical results slightly deviate from the analytic solution while maintaining the periodic oscillation behaviour. We conduct similar tests for other orbital configurations and find similar results; the secular solution describes the numerical results quite well as long as $\tau_\mathrm{acc}/P_\mathrm{orb}\gtrsim2$.

\section{Combined effect of mass-loss, natal kick and rocket}\label{sec:combination}

When \acp{sn} occur in binaries, the orbit will be altered by a combination of various effects. First, there is the immediate recoil due to the mass loss by \ac{sn} ejecta \citep[Blaauw mechanism;][]{bla61}, which causes the orbit to widen and change its eccentricity. Another effect is the instantaneous kick on the \ac{ns} associated with core collapse. The rocket mechanisms should then act on binaries that have already been perturbed by these rapid effects. The pre-rocket perturbations are crucial, as the rocket does not change the orbit unless the acceleration is tilted from the orbital angular momentum and/or there is initially non-zero eccentricity. Therefore, it is important to consider the full combined effect of the various forms of \ac{ns} acceleration.

\begin{figure}
 \centering
 \includegraphics[width=\linewidth]{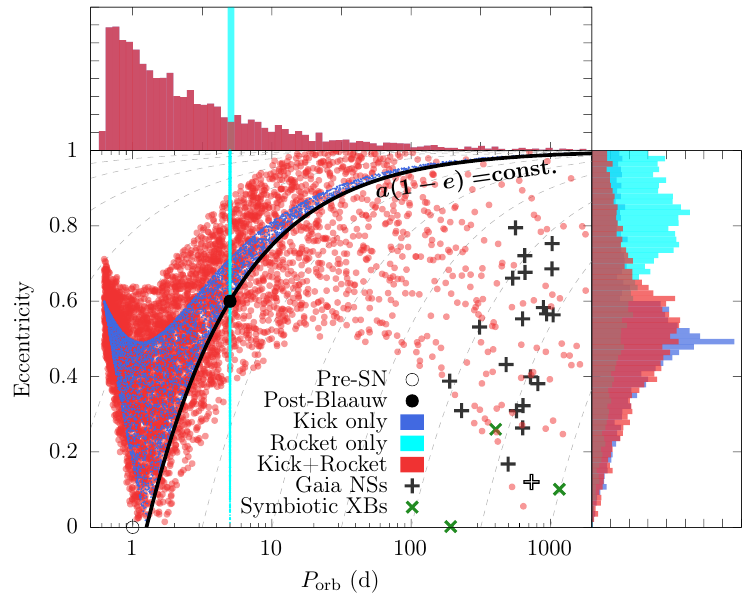}
 \caption{Period--eccentricity distributions of post-\ac{sn} binaries with different forms of \ac{ns} kicks. Grey dashed curves indicate fixed periastron distances and the black solid curve is where it equals the pre-\ac{sn} separation. Histograms on the top and right are the marginalized probability distributions of period and eccentricity, respectively. Black crosses ($+$) mark the locations of observed \ac{ns} systems discovered with Gaia \citep[]{elb24b}, whereas green crosses ($\times$) mark the location of symbiotic X-ray binaries with eccentricity measurements \citep[]{gal02,hin06,hin24,nag24}. Gaia NS1 is highlighted with an open cross.\label{fig:P-e_purekicks}}
\end{figure}

Figure~\ref{fig:P-e_purekicks} illustrates an example of how an initially circular orbit is altered by different combinations of rapid and rocket-like kicks. Initial system parameters are set to $m_1=3~\msun, m_2=1~\msun, P_\mathrm{orb,0}=1$~d, which has an average orbital velocity $\bar{v}_\mathrm{orb,0}\sim338$~\kms\footnote{Note that this is not the $\bar{v}_\mathrm{orb}$ relevant for the post-rocket orbit calculation.}. The \ac{sn} progenitor is assumed to have lost its hydrogen envelope through binary interactions and has become a helium star \citep[cf.][]{tau13}. We assume the primary star instantaneously loses $m_\mathrm{ej}=1.5~\msun$ in the \ac{sn}. After mass loss, the Blaauw effect causes the binary to jump both in period and eccentricity from the open circle to the filled circle. We further apply either an instantaneous kick with $\Delta v_\mathrm{kick}=0.5\,\bar{v}_\mathrm{orb,0}$ (blue dots), a rocket-like kick with $\Delta v_\mathrm{roc}=0.5\,\bar{v}_\mathrm{orb,0}$ (cyan dots), or a mixture of both where we apply a $\Delta v_\mathrm{roc}=0.1\,\bar{v}_\mathrm{orb,0}$ rocket kick after the $\Delta v_\mathrm{kick}=0.5\,\bar{v}_\mathrm{orb,0}$ rapid kick (red dots). For each model, we sample $10^4$ systems assuming an isotropic probability distribution for the kick/rocket direction. For the mixture model, we independently sample both the kick and rocket directions from isotropic distributions. For the kick-only model (blue dots), the distribution spreads out diagonally with a strong correlation between the post-\ac{sn} period and eccentricity. In the rocket-only model (cyan dots), the eccentricity is altered whereas the orbital period is unchanged. It is clear that the instantaneous and rocket-like kicks impact the orbit in qualitatively different ways.

In the kick+rocket mixture model (red dots), the period distribution is purely determined by the rapid component (see histograms on the top panel). The rocket effect then spreads the distribution out in the eccentricity direction. This enables the binary to cover a much wider area in period--eccentricity space from a given progenitor system, despite the rocket magnitude being rather small ($\Delta v_\mathrm{roc}=0.1\,\bar{v}_\mathrm{orb,0}\sim34$~\kms). As the rocket effect depends on the average orbital velocity (Eq.~(\ref{eq:hevo})), its effects are most pronounced in the longer period regions where it can allow the binary to reach low eccentricities, down to $e=0$.

\section{Implications}\label{sec:implications}

There are many implications of this wide phase space coverage by the kick+rocket mixture model, but in this paper we highlight a few select interesting applications.

\subsection{Wide low-eccentricity binaries}

An important distinct feature of the rocket effect is that the periastron distance can \textit{widen} after the \ac{sn} (see in Figure~\ref{fig:P-e_purekicks} how dots are distributed to the right of the black curve). Note that it is impossible to widen the periastron distance beyond the pre-\ac{sn} separation with only rapid kicks. This effect is enhanced when the rocket acts on orbits already strongly perturbed by the rapid kick because the systems that are kicked to wider orbits are more susceptible to the rocket effect and can be circularized more easily. Therefore, the kick+rocket combination can in principle widen the periastron distance by arbitrary amounts as long as the secular approximation holds in the rocket phase. 

Recent observations by the Gaia satellite have discovered a number of wide ($P_\mathrm{orb}\sim200$--$1000$~d) \ac{ns} binaries with $\sim1~\msun$ stellar companions \citep[]{elb24a,elb24b} (see black crosses in Figure~\ref{fig:P-e_purekicks})\footnote{Gaia is only sensitive to long-period systems. The majority of systems in Figure~\ref{fig:P-e_purekicks} would be the progenitors of low- or intermediate-mass X-ray binaries with much shorter orbital periods.}. Most notable is Gaia NS1, which has an orbital period of $P_\mathrm{orb}\sim731$~d and the lowest eccentricity $e\sim0.122$ \citep[]{elb24a}. These systems are considered to be progenitors for symbiotic X-ray binaries, which also tend to be on wide and low-eccentricity orbits, e.g. GX 1+4 ($P_\mathrm{orb}=1160.8\pm12.4$~d, $e=0.101\pm0.022$) \citep[]{hin06}, 4U 1700+24 ($P_\mathrm{orb}=404\pm3$~d, $e=0.26\pm0.15$) \citep[]{gal02} and IGR J16194-2810 ($P_\mathrm{orb}=192.5\pm0.2$~d, $e\lesssim0.018$) \citep{hin24,nag24}. It is challenging to devise formation channels for such systems in the traditional binary evolution framework, as the orbital periods are too short to avoid binary mass transfer from the \ac{sn} progenitor that will shrink their orbits down to much tighter periastron distances. Even if the post-mass-transfer separations can be kept wide enough, the rapid kick needs to be $\Delta v_\mathrm{kick}\lesssim30$~km~s$^{-1}$ in order to keep the eccentricity low enough, whereas typical observed kick velocities are $\Delta v_\mathrm{kick}\gtrsim100$~km~s$^{-1}$ \citep[]{hob05,ver17,igo20,wil21}. Dynamical assembly channels in dense cluster environments have also been ruled out due to their low eccentricity \citep[]{tan24}.

\begin{figure}
 \centering
 \includegraphics[width=\linewidth]{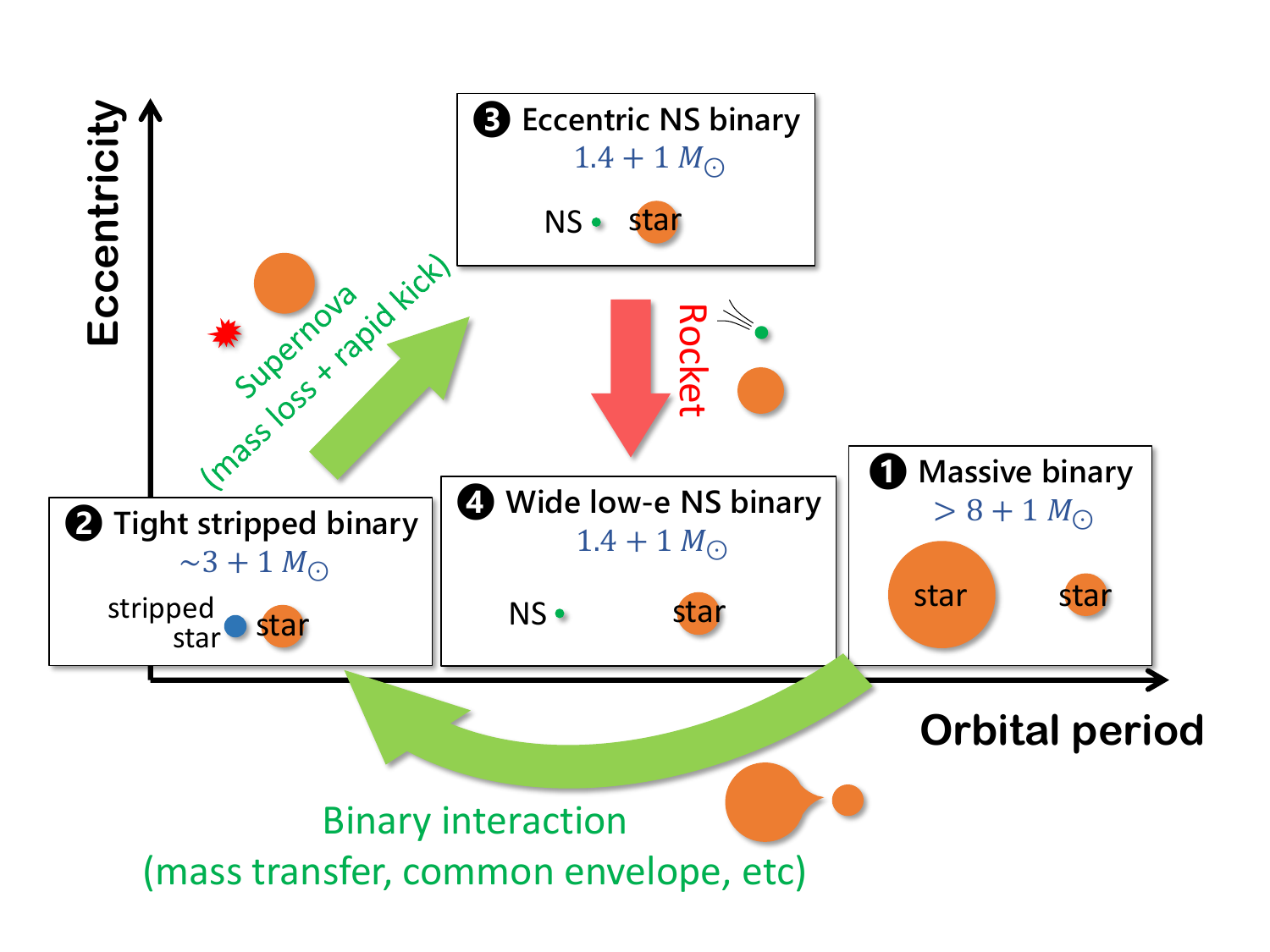}
 \caption{Schematic diagram of our formation channel for wide low-eccentricity systems like Gaia NS1. Each box (1--4) shows a specific phase of evolution and is placed in the period--eccentricity plane based on how it qualitatively evolves. \label{fig:schematic}}
\end{figure}

With rocket mechanisms, these systems can be formed more naturally (see Figure~\ref{fig:schematic}). The system can experience mass transfer or common envelope phases to strip the envelope of the primary and dramatically shrink the orbit. At this point the ejecta mass is small and the orbit is tight enough so that the \ac{sn} mass loss and natal kick tend to not disrupt the binary but instead send it on a wide and eccentric orbit. Once the orbit has widened, the average orbital velocity is now slow enough that even small rocket-like kicks ($\Delta v_\mathrm{roc}\lesssim30$~km~s$^{-1}$) can induce large changes to the eccentricity. 

In our demonstrative model in Figure~\ref{fig:P-e_purekicks}, the kick+rocket model (red dots) covers most of the period--eccentricity combinations of the observed systems. However, one of the observed symbiotic X-ray binaries (IGR J16194-2810) lies in the void at intermediate orbital periods where our model does not reach down to $e=0$. Such systems are difficult to form within the kick+rocket scenario alone \footnote{In order to form the circular orbit of IGR J16194-2810 within the kick+rocket scenario, it would require a rather large rocket velocity of $\Delta v_\mathrm{roc}\gtrsim50$~\kms.}. This indicates that an additional mechanism such as tidal circularization is required to explain the near-circular orbit of IGR J16194-2810. Given that the donor star is a giant and is inferred to be filling its Roche lobe by $\sim80~\%$ \citep[]{nag24}, it is very likely that tidal effects have efficiently circularized the orbit. Since tides are strongest at periastron, the orbit will evolve preserving the periastron distance during tidal interactions (along the grey dashed curves). Therefore, the orbit of IGR J16194-2810 could have originally been at longer periods and higher eccentricities, right in the range of the observed Gaia \ac{ns} systems. Conversely, the Gaia \ac{ns} systems may eventually all evolve to become circular-orbit symbiotic X-ray binaries like IGR J16194-2810, once the companion star evolves into giants and tidal effects become efficient.

Although calculating detailed formation rates of low-eccentricity \ac{ns} binaries is out of the scope of this letter, we note that classical binary evolution scenarios will need to be stretched to extremes whereas our scenario does not require any non-standard assumptions other than the rocket mechanism. Based on simple Monte Carlo simulations of kicks+rockets, we expect $\sim0.05$--$0.1~\%$ of \ac{ns} + $1~\msun$ star binaries to end up with wide ($P_\mathrm{orb}\geq100~\mathrm{d}$) and low-eccentricity ($e\leq0.3$) orbits if their pre-\ac{sn} periods are $1~\mathrm{d}\leq P_\mathrm{orb}\leq30~\mathrm{d}$. See Appendix~\ref{app:montecarlo} for details of the Monte Carlo calculations.

\subsection{Induced mergers}
Whereas some systems lower their eccentricities due to the rocket effect, some systems will \textit{increase} their eccentricities. If the rocket acceleration is directed at an appropriate angle, the eccentricity can reach very close to $e=1$, inevitably causing a tidal disruption of the companion or merger between the two components \citep{per16}. These interactions may explain the origin of some superluminous \acp{sn} or \acp{tzo} if the companion is a non-degenerate star \citep[]{mic16,RH22b}. 

In cases where the companion is also a compact object, it may induce a binary merger event that emits gravitational waves and/or electromagnetic signatures like short-duration \acp{grb}, kilonovae or other peculiar transients. The resulting transient will be separated from the \ac{sn} that gave birth to the \ac{ns} by a span of weeks to years, as the merger is induced by a secular effect. If, for example, a binary \ac{ns} merger occurs with a sufficiently short delay time, the merger will occur with unique conditions (e.g. relatively hot \ac{ns}, dense circumstellar material due to \ac{sn} ejecta), possibly giving rise to peculiar \acp{grb} or late-time bumps in the \ac{sn} light curve. Additionally if the delay time is long enough so that the \ac{grb} occurs after the peak of the associated \ac{sn} light curve, it may cause the accompanying \ac{sn} to be undetected like in some known long \acp{grb} \citep[]{fyn06,gal06,ras22,lev24}.

In prompt collisions induced by the instantaneous kicks, the fate is basically determined by the post-kick orbital configuration. With rockets, the orbit is not simply shifted to a different state but the system experiences a range of eccentricities and periastron distances over many orbital cycles so the probability of collisions can only increase. Therefore, we expect that the occurence rate of induced mergers would be higher than previous estimates based on classical kick models \citep[]{mic16,mic18}.

\subsection{Rockets in triples}
There are possibly even more interesting effects in the context of triple star evolution. For example if the \ac{sn} progenitor was in the inner binary of a hierarchical triple, the mass loss and natal kick not only alters the orbital period of the binary but also the outer orbit. In particular, the change in the outer orbit strongly depends on the orbital phase at which the explosion occurs, as it determines the direction of the mass-loss recoil. The subsequent rocket effect will cause the inner orbit to librate and change its eccentricity. As long as the \ac{ns} acceleration timescale is longer than the outer orbital period, the same libration will occur on the outer orbit too, as the inner binary as a whole will also be accelerated. It is not trivial which libration will be more significant. Both the inner and outer orbits have different initial configurations and angle with respect to the rocket. Also, the degree of libration and eccentricity change depends on $\Delta v_\mathrm{roc}/\bar{v}_\mathrm{orb}$, which is also different between the inner and outer orbits. The multitude of scales involved can lead to a wide variety of possible outcomes.

Once the kick and rocket effects have sufficiently tilted the inner and outer orbits, the mutual inclination may exceed the critical inclination required for secular evolution like Kozai-Lidov oscillations \citep[]{koz62,lid62} and higher-order effects \citep[]{nao16}. Or it could take the triple all the way to an unstable \citep[]{mar08,val08,tor22} or mildly unstable \citep{ant12,gri18} configuration which should eventually terminate through stellar mergers or ejection of one of the stars \citep[e.g.][]{per12}. Therefore, even if the rocket effect itself cannot modify the orbits enough to cause interesting phenomena, it can still be used as an efficient mechanism to hand over the system to where triple effects will finish off the job. Alternatively, the Kozai-Lidov oscillations may start to operate before the rocket effect has ceased. It may cause further interesting modulations and/or resonances if the oscillation timescales of the triple and rocket effects are comparable. We leave exploration of such phenomena for future work.

\section{Conclusion}\label{sec:conclusion}

Motivated by recent observations of non-axisymmetric magnetic fields on \acp{ns}, we explored the implications of combinations of rapid and rocket-like kicks in binary systems. We derive an analytic solution for the secular accelerated Kepler problem in vector form. This simple-formed solution is useful for other applications such as planetary science and artificial satellite maneuvering \citep[]{nam05,nam07,lan11}.

We demonstrate how the combined effect of mass loss, rapid natal kicks, and rocket-like kicks impact the post-\ac{sn} binary orbits. We find that even with only small contributions from the rocket component, the resulting binary orbit can be qualitatively different from what is achievable with purely rapid kicks. This can transform our understanding of the formation of many astronomical objects.

The wide low-eccentricity \ac{ns} binaries such as Gaia NS1 and GX 1+4 are challenging to form in classical binary evolution models. Our proposed new scenario based on rapid kick+rocket mixtures can more robustly form these systems without requiring drastic changes to how we understand binary interactions. We also expect that there could be many types of peculiar transients from mergers induced by the rocket effect between \acp{ns} and their companions. By studying these observables in detail, we may in turn gain insight into the physics that drive the rocket mechanism, such as the \ac{ns} birth spins, configuration of magnetic fields in new-born \acp{ns} or efficiency of cyclotron neutrino emission from superfluid vortices.

There are potentially numerous other applications where the rocket effect may be relevant, for example the formation of X-ray binaries, gravitational-wave sources, and triple star evolution. We plan to explore many of these other avenues in future studies.

\begin{acknowledgments}
The authors thank the anonymous referee for constructive comments that improved the manuscript.
RH thanks Ilya Mandel, Shoichi Yamada, Yongzhong Qian and Andrew Melatos for useful discussions during the early stages of this project. HN is supported by Grant-in-Aid for Scientific Research (23K03468) and also by the NINS International Research Exchange Support Program.
AH is supported by the Australian Research Council through grants DP240101786 and DP240103174.
\end{acknowledgments}

\vspace{5mm}

\software{gnuplot \citep[]{Gnuplot}
          }

\appendix

\section{Monte Carlo simulation of formation rates}\label{app:montecarlo}

Given that the magnitude of rocket-like kicks are still poorly constrained both observationally and theoretically, it is tricky to estimate formation rates of the various objects formed in our kick+rocket scenario. Here, we conduct a simple Monte Carlo simulation to provide ballpark estimates of the formation probabilities of Gaia NS1-like systems from given pre-\ac{sn} binary configurations. We define any system with $P_\mathrm{orb}\geq100~\mathrm{d}$ as ``wide'' and $e\leq0.3$ as ``low-eccentricity'' binaries. 

We first start from a system with a $3~\msun$ \ac{sn} progenitor and a $1~\msun$ companion star on a circular ($e=0$) orbit. We assume that $1.5~\msun$ is lost in the \ac{sn} and the primary now becomes a $1.5~\msun$ \ac{ns}. The progenitor is assumed to have undergone some form of binary interaction to lose a large fraction of its envelope so that the ejecta mass is within the typical range of observed stripped-envelope \ac{sn} ejecta masses \citep[]{lym16,tad18}. The Blaauw mechanism causes the binary to become wider and more eccentric after this. We then randomly draw the rapid kick velocity $\Delta v_\mathrm{kick}$ from a Maxwellian distribution with an average velocity $\sigma_\mathrm{kick}=265$~\kms \citep[]{hob05} and a random direction from a spherically isotropic distribution. After applying this rapid kick to the \ac{ns}, we check whether the binary is still bound and has an orbital period satisfying the wide condition ($P_\mathrm{orb}\geq100~\mathrm{d}$). If the period is long enough we then draw a random rocket velocity $\Delta v_\mathrm{roc}$ from a Maxwellian distribution with an average velocity $\sigma_\mathrm{roc}$ and a random direction from a spherically isotropic distribution. We then compute the post-rocket orbit and check if the resulting eccentricity satisfies the low-eccentricity condition ($e\leq0.3$).

Several models are computed with varying rocket velocity distributions ($\sigma_\mathrm{roc}=0, 30, 100~\mathrm{km}~\mathrm{s}^{-1}$) and initial orbital periods $P_\mathrm{orb,0}\in[0.1,1000]~\mathrm{d}$. For each model we draw $10^6$ samples to obtain the probability of forming wide and low-eccentricity systems. 

Firstly, we see that the fraction of systems that stay bound after the \ac{sn} decreases with increasing initial period (dashed). Among the surviving systems, only a fraction of them achieve long orbital periods satisfying the wide condition when the pre-\ac{sn} orbital period is not wide. Out of the total number of samples, the fraction of wide systems ranges $0.1$--$1~\%$ (dotted), peaking at around $P_\mathrm{orb,0}\sim4~\mathrm{d}$. 
When there is no subsequent rocket phase (solid black), it is basically impossible to form wide low-eccentricity systems unless the initial period is already long ($P_\mathrm{orb,0}\gtrsim50~\mathrm{d}$). On the other hand, with a rocket phase there is finite probability to form such binaries even if the pre-\ac{sn} period is significantly shorter than the threshold (solid red and pink). Among the wide binaries, the fraction that end up having low eccentricities only weakly depends on the initial period, ranging $\sim7$--$10~\%$.

\begin{figure}
 \centering
 \includegraphics[width=\linewidth]{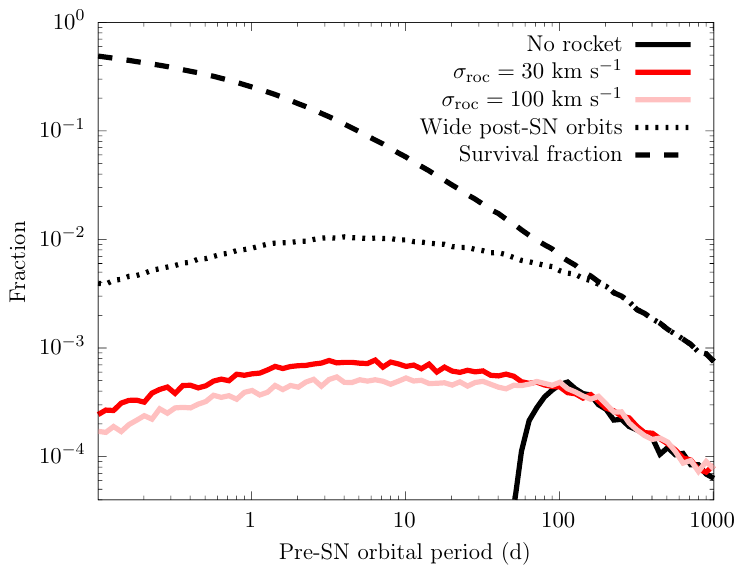}
 \caption{Fractions of systems that become wide ($P_\mathrm{orb}\geq100~\mathrm{d}$) and low-eccentricity ($e\leq0.3$) binaries. Different colours assume different average rocket velocities $\sigma_\mathrm{roc}$. The dashed curve shows the total survival fraction and the dotted curve shows the fraction of wide binaries regardless of the eccentricity.\label{fig:fractions}}
\end{figure}

Our main conclusion is that as long as the binary can achieve long orbital periods after rapid kicks, there is almost a fixed probability ($\sim7$--$10~\%$) of forming low-eccentricity binaries through the rocket mechanism as long as the rocket magnitude is comparable to the post-kick orbital velocity. This is in contrast to pure-kick models where it is impossible to create wide low-eccentricity binaries unless the pre-\ac{sn} orbit is already wide. Orbital periods around $P_\mathrm{orb,0}\sim100~\mathrm{d}$ is still too short to avoid any binary interactions between the massive \ac{sn} progenitor and the low mass companion. Typically, such interactions would tighten the orbit down to much shorter orbital periods. Therefore, if we integrate our distribution in Figure~\ref{fig:fractions} weighted by the distribution of post-binary-interaction periods, pure-kick models have a near-zero chance of forming wide low-eccentricity systems, especially Gaia NS1. Given that dynamical assembly in clustered environments have been proven to be difficult to create Gaia NS1-like systems \citep[]{tan24}, triple evolution or our kick+rocket scenario may be the only plausible explanations for the formation of Gaia NS1-like binaries.

In this simple analysis we have ignored various possible effects that may influence the probabilities. For example, we have assumed that the kick, rocket and orbit are independent whereas there may be correlations between the magnitude and direction of these components \citep[]{bur24}. There may be correlations between the pre-\ac{sn} orbital periods and the rocket magnitudes due to tidal spin-up of the \ac{sn} progenitor \citep{ful22}. Different forms of kick velocity distributions may impact the overall survival and wide-orbit fractions. The thresholds we use for ``wide'' and ``low-eccentricity'' are arbitrary choices. We leave detailed exploration of formation rates of Gaia NS1-like binaries to future work.

\bibliographystyle{aasjournal}

\end{CJK*}
\end{document}